# Microelectrode arrays of diamond-insulated graphitic channels for real-time detection of exocytotic events from cultured chromaffin cells and slices of adrenal glands

Federico Picollo[1,2,3,4], Alfio Battiato[2,1,3,4], Ettore Bernardi[2,1,3,4], Andrea Marcantoni[5,3,4], Alberto Pasquarelli[6], Emilio Carbone[5,3,4], Paolo Olivero[2,1,3,4], Valentina Carabelli[5,3,4,*]

[1]Istituto Nazionale di Fisica Nucleare (INFN), sez. Torino, Italy.
[2]Physics Department, University of Torino, Italy.
[3]"Nanostructured Interfaces and Surfaces" Inter-departmental Centre, University of Torino, Torino, Italy.
[4]Consorzio Nazionale Interuniversitario per le Scienze fisiche della Materia (CNISM), Torino Unit, Italy.
[5]Department of Drug Science and Technology, University of Torino, Italy.
[6]Institute of Electron Devices and Circuits, Ulm University, Ulm, Germany.

\* Corresponding author:   valentina.carabelli@unito.it
Fax number: 0039 011 6708498

# ABSTRACT

A microstructured graphitic 4×4 multielectrode array was embedded in a single-crystal diamond substrate (4×4 μG-SCD MEA) for real-time monitoring of exocytotic events from cultured chromaffin cells and adrenal slices. The current approach relies on the development of a parallel ion beam lithographic technique, which assures the time-effective fabrication of extended arrays with reproducible electrode dimensions.

The reported device is suitable for performing amperometric and voltammetric recordings with high sensitivity and temporal resolution, by simultaneously acquiring data from 16 rectangularly shaped microelectrodes (20×3.5 $\mu m^2$) separated by 200 μm gaps. Taking advantage of the array geometry we addressed the following specific issues: *i)* detect both the spontaneous and KCl-evoked secretion simultaneously from several chromaffin cells directly cultured on the device surface, *ii)* resolve the waveform of different subsets of exocytotic events, *iii)* monitoring quantal secretory events from thin slices of the adrenal gland.

The frequency of spontaneous release was low (0.12 Hz and 0.3 Hz respectively for adrenal slices and cultured cells) and increased up to 0.9 Hz after stimulation with 30 mM KCl in cultured cells. The spike amplitude as well as rise and decay time were comparable with those measured by carbon fiber microelectrodes and allowed to identify three different subsets of secretory events associated to "full fusion" events, "kiss-and-run" and "kiss-and-stay" exocytosis, confirming that the device has adequate sensitivity and time resolution for real-time recordings. The device offers the significant advantage of shortening the time to collect data by allowing simultaneous recordings from cell populations either in primary cell cultures or in intact tissues.

# INTRODUCTION

Single-cell amperometry allows the detection of quantal fusion events with sub-millisecond time resolution and picoampere sensitivity [1,2]. In the last decade, conventional approaches using carbon fiber electrodes have been combined to a variety of chip-based planar arrays [3,4], with the aim of increasing the spatial resolution and providing an electrochemical mapping of exocytosis within a single cell [5-7], or collecting events from multiple samples [8-14]. For a recent comprehensive review on the subject, see [15,16]. Focusing on multiple detection of vesicular release, arrays with up to 64 or even 10×10 electrodes with subcellular dimensions have been produced [12,13], thus allowing a multi-site detection from chromaffin, PC-12 cells, striatal slices [17-20] and the coupling of amperometry with fluorescence microscopy [21,22]. An appealing substrate material for these applications is diamond, which offers a wide spectrum of physical-chemical properties which are crucial for the realization of integrated planar sensors, namely: wide optical transparency from infrared (IR) to near ultraviolet (NUV), high chemical inertness [23], good biocompatibility [24] and the possibility of directly writing sub-superficial graphitic microelectrodes by means of MeV ion beam lithography [25-31].

Taking advantage of the above-mentioned properties, several prototypes of diamond-based electrochemical sensors with integrated graphitic microelectrodes have been developed and tested. A first prototype allowed to detect quantal secretion from individual chromaffin cells positioned over a single microelectrode using a patch-clamp glass pipette [32]. The device was subsequently developed into a multi-electrode array consisting of several sensing regions over the same substrate, allowing the quantification of oxidizable analytes dispersed in aqueous solution [33]. Due to the interest in the electrochemical mapping of exocytosis from single cells, a new sub-micrometric fabrication technique was also developed for constructing high-density multielectrode arrays able to resolve micro-domains of secretion [6,34].

In the present work, we adopted a parallel ion beam lithographic technique, which allows the time-effective fabrication of microelectrode arrays in diamond substrates with better reproducible dimensions with respect to the previous fabrication strategy. To our knowledge, this represents the first attempt, using diamond-insulated micrographitic channels, to provide a multi-site detector for measuring exocytosis from cultured cells or tissue slices.

From a biological point of view, the 4×4 micro-Graphitic Single Crystal Diamond Multi Electrode Array (4×4 μG-SCD MEA) has the advantage of allowing to culture living cells for several days and to measure catecholamine secretion without electrode fouling up to five days since plating, while the array geometry (16 channels patterned within a 600×600 $\mu m^2$ area) speeds up the amperometric detection from multiple samples. These features make the chip suitable for addressing relevant physiological issues related to the detection of spontaneous and stimulated catecholamine secretion, as well as the discrimination between events associated with complete ("full fusion") and incomplete vesicle fusion ("kiss-and-run" and "kiss-and-stay" exocytosis) [35]. In particular, the relevance of spontaneous and evoked secretion is related to the *in vivo* supply of catecholamines from the adrenal gland that has to be continuously tuned on the basis of the hormonal need, thus ranging from a basal resting release to a more sustained secretion triggered by continuous cell stimulation mimicking stressful conditions.

Cultured mouse chromaffin cells (MCCs) exhibit spontaneous firing activity [36] finely modulated by the coupling of calcium and calcium-dependent potassium channels [37,38] and the availability of sodium channels [39]. Taken together, these mechanisms contribute to regulate catecholamines release [40] that occurs spontaneously both in isolated MCCs and mouse adrenal slices. In the latter, secretion occurs in cell clusters [41] and recorded by conventional CFEs [42]. In the present work we exploited the 4×4 µG-SCD MEA to detect and characterize quantal secretory events, under resting conditions and during cell depolarization, both from isolated cells and tissue slices. We used bovine chromaffin cells (BCCs) to ensure an adequate cell density on the electrodes when detecting exocytosis from isolated samples, whereas tissue slices were obtained from mouse adrenal glands. The use of intact adrenal tissue represents an undisputed challenge to mimic physiological conditions. This has been initially approached by stimulating the whole adrenal gland by electric shock or by perfusing blood vessels to induce massive catecholamine release [43]. Subsequently, the resolution of fast secretory responses was characterized by monitoring capacitance increases from adrenal slices [44-46]. More recently, spontaneous and stimulated secretion was monitored from mouse adrenal slices by means of conventional carbon fibers [42,41]. As shown, the chip is extremely effective in resolving single secretory events of extremely low amplitude (8 pA) and fast rising phase (1 ms) simultaneously from many cells, thus resulting in a powerful tool for *in-situ* biological investigation of neurosecretion in living cells.

**EXPERIMENTAL**

The device employed in this study consists of 16 sub-superficial nanocrystalline graphitic micro channels converging into a central region where their endpoints (i.e. the sensing electrodes) are exposed to the sample surface. Each sensing electrode (area of the channel exposed to the solution) has a ~20×3.5 µm$^2$ surface area, and the 16 electrodes are arranged on a 4 × 4 square grid with 200 µm spacing. At the substrate periphery, the other emerging endpoints of the micro-channels provide electrical contacts for bonding to a dedicated chip carrier. Further details on the realization process and final configuration of the device are reported in Supplementary Information and shown in Figure 1S and 2S.

**RESULTS AND DISCUSSION**

*Electrical characterization of the 4×4 µG-SCD MEAs*

In a preliminary series of experiments, we first characterized the conduction and electrochemical properties of the sub-superficial graphitic microchannels. To determine their electrical conductivity, we recorded two-terminal current-voltage (I-V) characteristics by means of microprobes connected with suitably shielded feedthroughs to a Keithley 2636 electrometer. Two metal pads were temporarily deposited at the end-points of each channel in order to guarantee a low contact resistance (see inset in Figure 1A). The I-V characteristics display ohmic behavior with resistance between 5.3 kΩ and 7.7 kΩ, depending upon the microelectrode length (Figure 1A). Also the parasitic surface resistance between nearby graphitic electrodes was measured, confirming the highly insulating property of the single-crystal diamond matrix. There was no detectable electrical conductivity within the instrumental limits, corresponding to insulating resistance values > 1 TΩ. The resistivity of each conductive channel was evaluated from its resistance and geometrical dimensions, obtaining a mean value of ρ = 2.1 ± 0.3 mΩ cm, comparable with that of standard polycrystalline graphite, i.e. ~1.3 mΩ cm [47].

The electrochemical properties and the capacitance of the channels were determined by cyclic voltammograms simultaneously recorded from the 16 microelectrodes. The measurements were performed with a home-developed setup by sweeping the voltage within the electrochemical window of the graphitic electrodes, i.e. from –0.5 V to +1.2 V. The voltage was applied with respect to a Ag/AgCl reference/counter electrode at different scan-rates ranging from 10 mV s$^{-1}$ to 400 mV s$^{-1}$ (Figure 1B). To assess the capability of our device to detect the adrenaline oxidation peak and to confirm the micrometric dimension of electrode sensitive areas we used a Tyrode solution with 1 mM adrenaline for all the experiments (see Supporting Information).

In Figure 1B three different features can be clearly recognized in each scan performed at a different rate: i) a region ranging from –0.5 V to +0.4 V with negligible redox currents, ii) a high current peak at +1.2 V corresponding to water hydrolysis, and iii) a shoulder associated with adrenaline oxidation centered around +0.8 V. All the microelectrodes exhibit similar cyclic voltammograms. The optimal bias voltage for the subsequent amperometric measurement was therefore set to +0.8 V, in correspondence of the maximum value of the ratio between the adrenaline oxidation current and the water hydrolysis background current.

Moreover, it is worth noting that all the voltammograms acquired at different scan rates show similar waveforms. This feature can only be ascribed to the micrometric extension of the active surface area of the graphitic electrodes [48].

The evaluation of the double layer specific capacitance ($C_{dl}$) was performed by using the following equation [49]:

$$\quad - \quad —$$

where $S = 70$ µm$^2$ is the area of the surface electrode, $\Delta i$ is the difference between the cathodic and anodic currents at the open circuit potential and $v$ is the voltage scan rate. Figure 1C shows the voltage region where $\Delta i$ was evaluated. Consistently with the above equation, $\Delta i$ displays a linear dependence from $v$, as shown in Figure 1D. From a linear fitting of the "$\Delta i$ vs $v$" data (Figure 1C), and upon normalization on the above-mentioned surface area, a $C_{dl} = (2.24 \pm 0.09)$ mF cm$^{-2}$ value with 20 % of RSD overall the 16 channels was obtained.

Furthermore, even though the device was primarily developed to detect exocytosis from chromaffin cells, where adrenaline stored into vesicles reaches concentration of ~0.5 M, its response to adrenaline (Figure 3S) is lower than µM, since it has Limit of Detection (LOD) of 0.37 µM and Limit of Quantification (LOQ assumed as equal to 3×LOD) of 0.74 µM. The lower limit of quantification guarantees the sensor applicability also on biological system with lower amount of neurotransmitter release (i.e. neuronal network, etc.).

*Spontaneous catecholamine secretion from BCCs*

Cultured chromaffin cells exhibit spontaneous firing activity: action potential discharges are triggered by the onset of a pacemaker current and regulate catecholamine release [50]. Spontaneous firing of action potentials occurs at 1.5 Hz frequency in MCCs and is driven by a tight coupling of L-type calcium channels and Ca$^{2+}$ dependent potassium channels [51,52]. Beyond this well-established coupling between $Ca^{2+}$ ad $K^+$ channels, an additional regulation of action potential shape, burst firing and catecholamine release may occur on the basis of $Na^+$ channels availability: steady depolarization reducing $Na^+$ channels availability switches the firing mode from tonic to bursts, leading to increased catecholamine secretion [39]. Although not yet proved, it is likely that the same occurs in BCCs, which are chromaffin cells easier to isolate and culture. They are available at very large quantities and, thus, can be plated at high densities to cover most but not all the recording electrodes.

BCCs were plated on the 4×4 μG-SCD MEA at a density of approximately 150,000 cells per chip, to test whether spontaneous secretory activity could be detected under physiological conditions (external standard Tyrode solution containing 2 mM $Ca^{2+}$). For sake of consistency with previous studies, experiments were generally performed within 3 days after plating; though, in order to test the responsiveness of the microarray versus time, amperometric detection was carried on up to five days with no detectable decrease in the electrode sensitivity. Amperometric recordings were performed, under sterile conditions, from the same cultured cells whose secretory activity was detected at different days (1, 3 and 5 days after plating) exhibiting the same spike parameters (Table S1, Figure 4S). Culturing chromaffin cells on diamond-insulated micrographitic arrays for longer periods is possible, due the biocompatibility of the substrate, though this would imply to induce a neuronal phenotype, exhibiting different properties of exocytosis[53], which goes far beyond the purposes of this work. The possibility of culturing cells on the device thus represents a relevant improvement with respect to previous prototypes [6,9,32] where the chromaffin cell was positioned one by one onto the electrodes immediately prior the recording to avoid electrode fouling, thus severely limiting the simultaneous acquisition from different cells.

With the present 4×4 μG-SCD MEA, we found that under control conditions secretion from BCCs occurred spontaneously in approximately 10% of cells with a mean frequency of 0.3 Hz. Figure 2 shows that in 3 out of 16 electrodes the spontaneous catecholamine secretion was monitored for several minutes and then blocked by adding 200 μM $CdCl_2$ to the external solution. The release frequency strictly depends on cell firing mode. In current-clamped MCCs using CFEs, catecholamine release occurs at a rate of 0.2 Hz when associated to tonic action potential firing, and increases up to 0.5 Hz when cell activity is switched to burst firing [39,54]. The rate of release detected by our device in BCCs (0.3 Hz) is intermediate between these two values. Given that the electrodes active surface of the 4×4 μG-SCD MEA (70 μm$^2$) approximately corresponds to that of CFEs[6], we can conclude that the two recording systems give comparable frequency values of secretory events [55].

*KCl-evoked catecholamine release in BCCs*

The 4×4 μG-SCD MEA was subsequently tested for recording the exocytosis from multiple samples following perfusion with high-KCl solutions to induce sustained secretory activity. Monitoring of amperometric spikes from many cells in an integrated device is an important task for speeding up the screening process of exocytotic events with respect to the CFEs. Figure 3A shows that simultaneous recordings from 5 out of 16 electrodes is possible with the 4×4 μG-SCD MEA. Cells were plated without any adhesion promoter (see Supporting Information). Interestingly, even before KCl stimulation, in saline solution containing 10 mM $Ca^{2+}$ spontaneous activity is detected with a mean frequency of $0.080 \pm 0.004$ Hz. The cumulative charge of spontaneous release, measured over 120 seconds was on average $14.6 \pm 3.8$ pC. On average, approximately 10% of the electrodes exhibited spontaneous activity, whereas no activity was present in the remaining channels. It is worth remarking that the above-mentioned spontaneous rate of release monitored in 10 mM $Ca^{2+}$ is significantly lower than the one detected in 2 mM physiological $Ca^{2+}$ (0.3 Hz). A reason for this is that catecholamine release is strictly dependent on cell electrical activity. Since the tight coupling between $Ca^{2+}$ and SK/BK channels mainly controls the action potential shape and firing frequency [39,51,52], it is likely that the increased $Ca^{2+}$ concentration leads to an enhanced activation of BK and SK channels that causes sustained cell hyperpolarization and consequent lowered rate of catecholamine release.

Cell stimulation with an external saline solution enriched with 30 mM KCl (Figure 3A, for the whole duration of the experiment) increased the spike frequency, ranging from 0.2 Hz to 1.9 Hz, with a mean value of $0.9 \pm 0.4$ Hz. The cumulative charge of KCl-evoked release was evaluated over 120 seconds, resulting in an average value of $172 \pm 14$ pC. The increase of this value with respects to the one measured during spontaneous secretion is uniquely ascribed to a more sustained frequency of release, since the spike parameters are not significantly altered (Table S1). Two examples of amperometric spikes detected from electrode 1 are zoomed in Figure 4B. KCl-induced exocytosis was completely blocked by adding 200 μM $CdCl_2$ to the external solution (not shown here), to exclude the recording of artifacts. Direct comparison with previous reported data is complicated by the rather heterogeneous experimental approaches used by several groups (substrate, cell adhesion, stimulation protocols, detection of exocytosis from cell apex or cell bottom). Limiting the comparison to BCCs, a mean frequency of 0.008 Hz μm$^{-2}$ was found from cells adherent on ITO electrodes and stimulated with external $BaCl_2$ [55], whereas KCl induced a mean frequency of 0.011 Hz μm$^{-2}$ in cells positioned on boron-doped ultra-microelectrode arrays [6]. Assuming that the whole electrode surface of our chip (~70 $\mu m^2$) was in contact with the cell, we estimate a mean frequency of ~0.013 Hz $\mu m^{-2}$. This slightly higher value may derive from the way BCCs were plated and cultured on the 4×4 µG-SCD MEA, i.e. at high density and avoiding any kind of cell manipulation, whereas in the previous cases the isolated cell was positioned onto the chip just before the experiment. Cell mechanical manipulation and paracrine modulation from neighboring cells are likely to affect spontaneous firing and rate of catecholamine release.

In order to validate the device sensitivity, spike parameters were compared with those obtained using conventional CFEs with BCCs cultured on the same 4×4 µG-SCD MEA substrate. As shown in Table S1, the measured spike parameters were compatible among the indicated experimental conditions and in good agreement with the reported values on BCCs [55,56].

*Detection of different modes of exocytosis in BCCs*

Different populations of amperometric spikes have been identified either in MCCs or BCCs by means of conventional carbon fibers, 4×4 µG-SCD MEAs and boron-doped diamond MEAs [6,32,36,57]. In these studies, the $Q^{1/3}$ values were bimodally distributed accordingly to a double Gaussian function, suggesting the existence of two populations of vesicles or two different modes of secretion. However, a detailed analysis of spike parameter distributions both in MCCs and PC12 cells uncovered three distinct modes of exocytosis [35,58,59]: "full fusion" events (large events), "kiss-and-run" events (small events with fast decaying kinetics) and "kiss-and-stay" associated to "stand alone foot" events (SAF). In this latter mode, the fusion pore reverses and closes without achieving full fusion giving rise to small unitary currents with slow decay. Our goal here was to test the capability of the 4×4 µG-SCD MEA to reveal the presence of low-amplitude SAF events, which were overlooked in previous analysis [32].

In order to investigate the presence of different clusters of events, we first used 2D Gaussian mixture analysis of spike parameters [35]. Figure 4A reports the scatter plot of Log($Q$(fC)) versus Log($I_{max}$(pA)) values, where variances are represented by solid curves. Three 2D Gaussian distributions uncovered the existence of different clusters of large, small and SAF events, in analogy with[35]. For all the events we found direct proportionality between Log($Q$(fC)) and Log($I_{max}$(pA)), the fitted slopes being 1.2, 1.3 and 1.5 for large, small and SAF events, with corresponding correlation coefficients of 0.86, 0.91, 0.71 respectively. More specifically, large, small and SAF events contributed respectively to 46%, 42% and 11% of the total number of recordings. For SAF events, log-transformed distributions of charge, maximum current amplitude and decay time are plotted separately and fitted by 1D Gaussian functions (Figure 4B). Considering the mean values of the 2D mixture analysis corresponding to the center of the ellipsoids, the Log($I_{max}$(pA)) values are in good agreement with the means of 1D distributions, as shown in figure 4B: 2.3 in both cases for large events, 1.7 versus 1.6 for small events, and 1.2 versus 1.1 for SAF events. A similar correspondence was found for Log($Q$(fC)) values: 3.3 versus 3.4 for large events, 3.0 in both cases for small events, and 2.9 in both cases for SAF events. SAF events were characterized by rise times ($\tau_{rise}$) of <3 ms and thus were likely originated close to the electrode. They were also characterized by small amplitudes (<20 pA) and slowly decaying currents (i.e. tens of ms, Figure 4C).

It is worth pointing out that SAF events do exhibit a significantly reduced quantal charge respect to small and large events, as well as a prolonged mean decay time $t_{decay} = 42 \pm 7$ ms, indicating much slower decay kinetics. This evidence confirms what was previously reported [35]. The 790 fC charge estimated for the SAF events of BCCs is significantly higher than what reported for MCCs by Mackenzie et al. [59], i.e. 120 fC. This may be partly ascribed to differences of the catecholamine content in the granules of MCCs (approximately 3-fold, [6]), and also to a larger peak-to-peak noise, which may underestimate the number of smallest events. This assumption is also supported by the observation that the mean charge evaluated for all exocytotic events grouped together (1.2 pC) is in very good agreement with published data on BCCs [55,56].

*Quantal events detected from mouse adrenal slices*

Aside from investigating the process of exocytosis from isolated cells, here we tested the capability of the 4×4 μG-SCD MEA to detect catecholamine release using adrenal medulla slices, to preserve most features of the intact adrenal gland.

Due to the large dimensions of bovine glands, and for a better comparison with previous data, amperometric recordings from adrenal slices were performed using mouse adrenal glands. Figure 5 shows spontaneous events monitored simultaneously from 5 different channels in the presence of 2 mM $Ca^{2+}$ (SOL 2, Supplementary Information). Nevertheless a higher number of cells was exposed over the device respect to a chromaffin cells culture, a reduced signal frequency could be observed (p<0.05), presumably because of the different culture conditions. Spike frequency of spontaneous secretion was $0.12 \pm 0.04$ Hz and was suppressed by superfusing the tissue either with a low $Ca^{2+}$ containing solution for 3-5 minutes (SOL 1, Supplementary Information) (Figure 5), or by holding the electrodes polarization at 0 mV (not shown here). The spike parameters estimated from these experiments are reported in Table S1 (Supplementary

Information). They closely resemble those previously reported on mouse adrenal slices [41,42] even though these latter were obtained using different stimulation protocols (KCl, acetylcholine, electrical depolarization). Validation with tissue slices further increases the range of application of micrographitic arrays toward more physiological conditions.

## CONCLUSIONS

The 4×4 µG-SCD MEA is shown to be an extremely versatile device suitable for real-time monitoring of both spontaneous and stimulated exocytosis. The excellent biocompatibility of the substrate allows stable recordings from long-term primary cultures. This device, in our knowledge, represent the first diamond-based sensor that allows directly culture cells over its surface for several days performing amperometric measurements simultaneously from isolated cells. Moreover, although the employment of graphitic electrodes could represent detriment of the electrochemical properties respect to nanocrystalline diamond electrodes, the reported fabrication process offers the advantage of avoiding the introduction of passivation layers during the devices realization. This represents a main strength in terms of mechanical/chemical stability and lifetime (high number of repeated measurement cycles). Indeed, the deviced operated with unchanged performance after several measurement sessions spread out over more than 12 months. Besides these features, the high time resolution and sensitivity allow the clear separation of three differently shaped amperometric events: "full fusion", "kiss-and-run" and "kiss-and-stay" exocytotic events. This detailed analysis is unprecedented on chip-based planar arrays. Moreover, the employment of the graphite as electrode material allows to acquire data compatible with standard CFEs, allowing the direct comparison with the literature, while still exploiting the significant advantages offered by the employment of a diamond substrate (robustness, transparency, biocompatibility, chemical inertness and resistance to biofouling).

Our work demonstrates the potentiality of our all-carbon biosensor as an attractive advanced device for investigating cell functioning in intact tissues of neuroendocrine glands and complex neuronal networks during synapses maturation in brain slices. Due to the simultaneous recording from multiple cells, the device is particularly attractive for drug screening tests on various cell types in culture, possibly in combination with optical fluorescence measurements, thus exploiting the excellent optical transparency of diamond. Moreover, the process of fabrication of graphitic microchannels is suitable for coupling the electrochemical measurements with microfluidic channels that can be obtained via the selective removal of the graphite [60]. This feature would envisage an entirely new class of devices, with significant advantages with respect to the state of the art.

## ASSOCIATED CONTENT

**Supporting information**

It provides details of the *4×4 µG-SCD MEA* fabrication, methods related to culture and isolation procedures of chromaffin cells and adrenal slices, amperometric detection of exocytosis and data analysis. Table S1 includes amperometric spikes parameters under different conditions and days in culture, and Figure 1S is related on the fabrication process of the device. Figure 2S shows a 3D scheme of the array, Figure 3S the dose-response curve for adrenaline, Figure 4S some representative traces of amperometric recordings at different days in culture.

## ACKNOWLEDGMENTS

This work is supported by the following projects: "DiNaMo" (young researcher grant, project n° 157660) by the Italian National Institute of Nuclear Physics; FIRB "Futuro in Ricerca 2010" project (CUP code: D11J11000450001) funded by the Italian Ministry for Teaching, University and Research (MIUR); "A.Di.N-Tech." project (CUP code: D15E13000130003) funded by the University of Torino and "Compagnia di San Paolo" and Italian MIUR (PRIN 2010/2011 project 2010JFYFY2 ), Torino University Local Grants to VC, AM and EC. The MeV ion beam irradiation was performed at the "AN2000" Van-der-Graff accelerator of the INFN Legnaro National Laboratories (INFN-LNL), within the "Dia.Fab." experiment.

**Figure 1.**

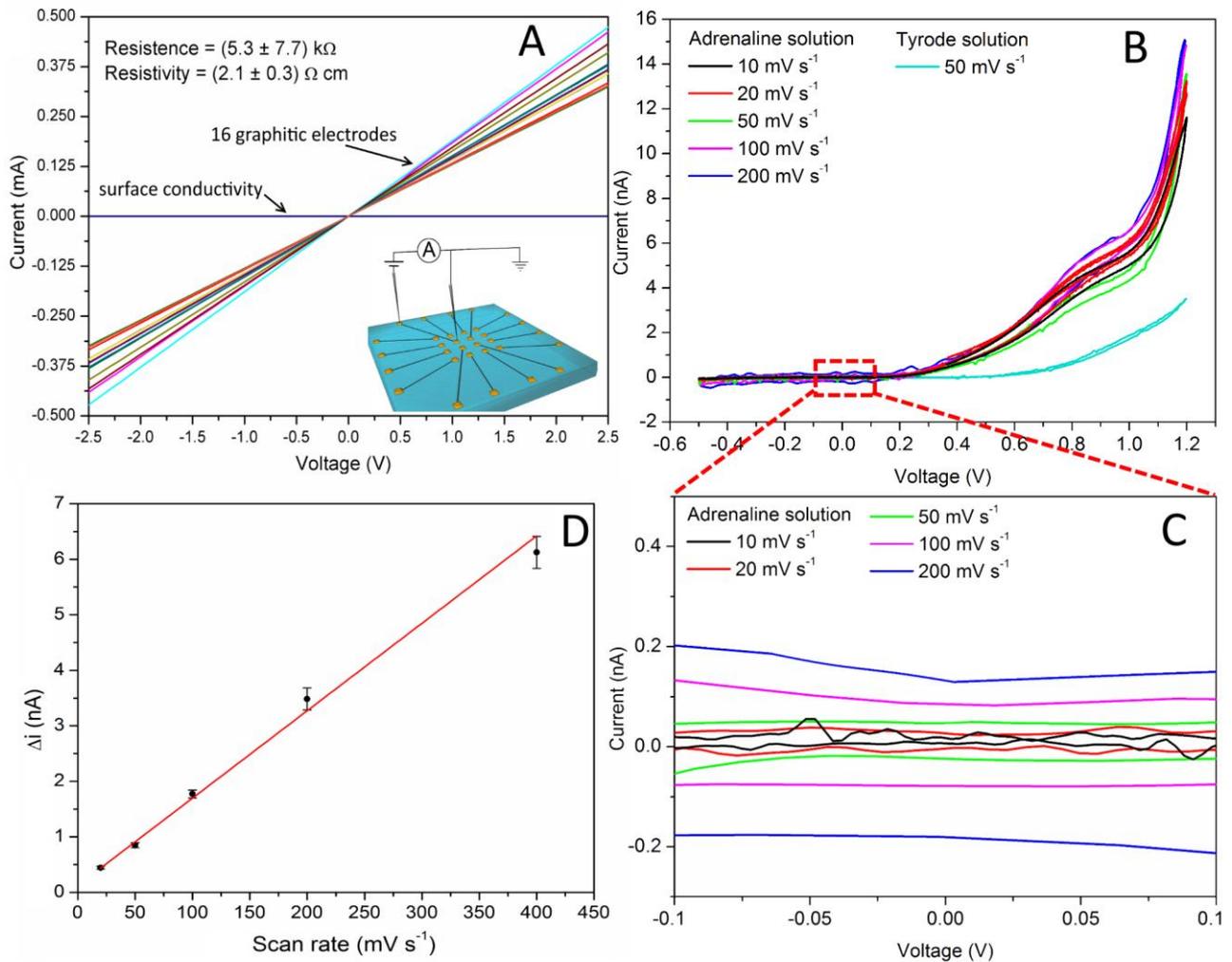

**Figure 1. 4×4 µG-SCD MEA electrical characteristics**

A) Current-voltage characteristics of the 16 graphitic electrodes (displaying ohmic behavior) and of the parasitic surface conduction path (dark blue). B) Steady-state voltammograms of Tyrode + Adrenaline solution (1 mM) and Tyrode solution acquired at different scan rates (from 10 mV s$^{-1}$ to 200 mV s$^{-1}$); comparison with the curves obtained in adrenaline containing solutions highlights the presence of the "shoulder" associated with adrenaline oxidation, approximately centered around +0.8 V. C) Magnification of the cyclic voltammetry measurement at open circuit potential: $\Delta i$ current was evaluate in this region. D) $\Delta i$ vs. scan rate plot: linear fit was performed in order to evaluate the electrodes capacitance.

**Figure 2.**

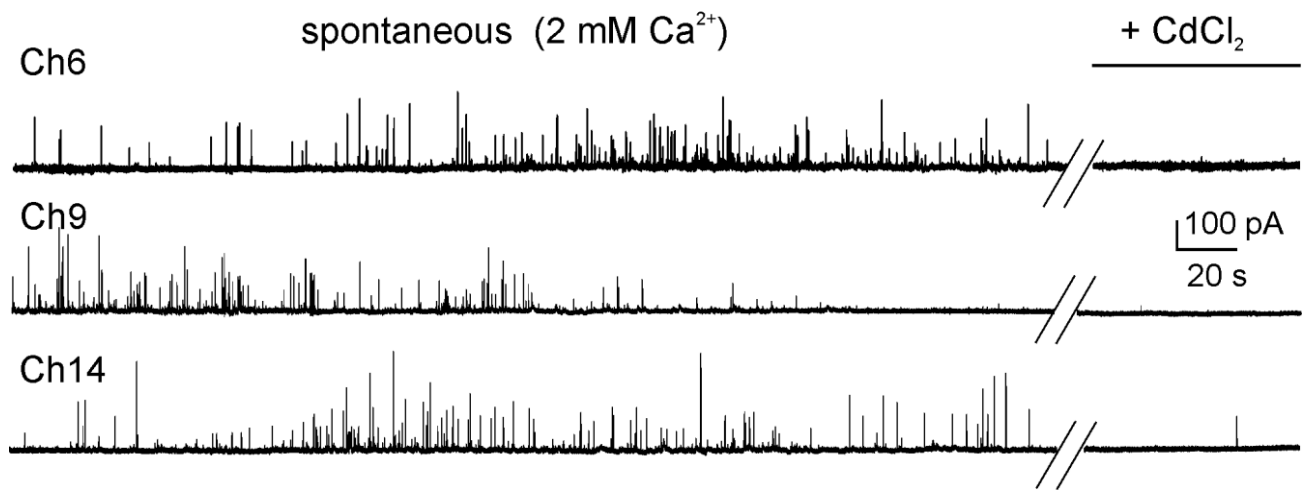

**Figure 2. Spontaneous exocytosis from isolated BCCs**

Representative traces showing spontaneous amperometric spikes recorded from isolated BCCs bathed in 2 mM $Ca^{2+}$. After the interruption (//) the secretory activity is suppressed by applying 200 µM $CdCl_2$ to the bath (horizontal line).

**Figure 3.**

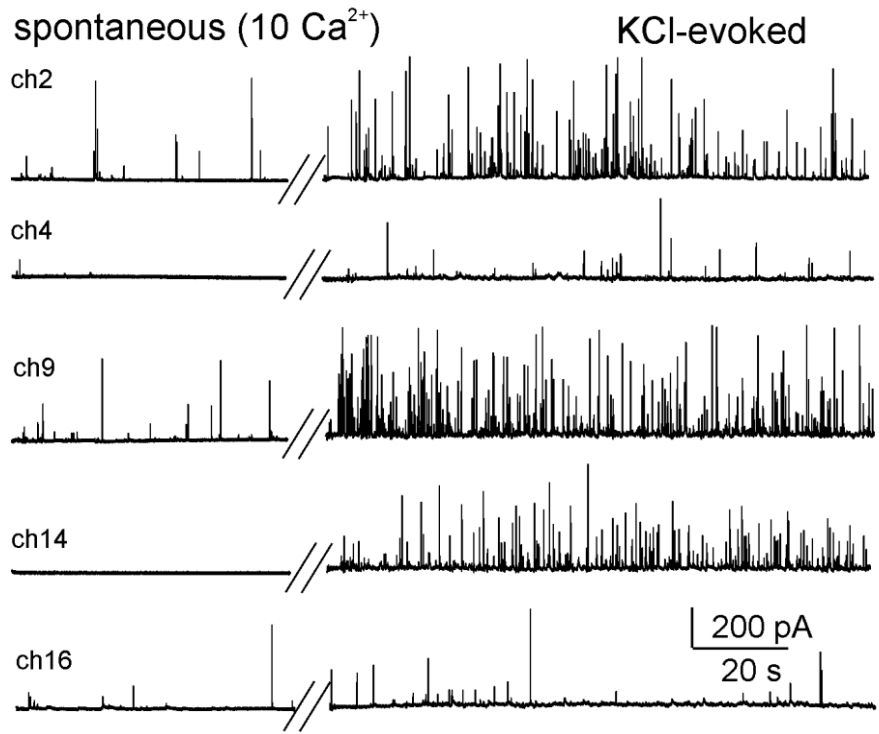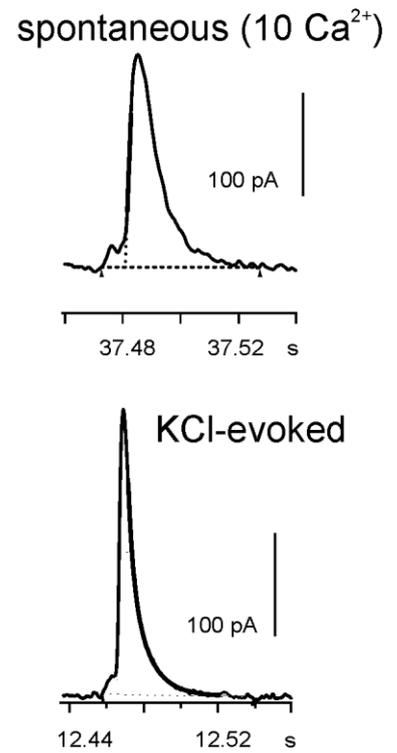

**Figure 3. KCl-induced increase of the rate of exocytosis**

A) Representative traces from BCCs bathed in Tyrode solution containing 10 mM $Ca^{2+}$ (left) and then perfused with a KCl-enriched solution. B) Examples of single spikes obtained before and after stimulation with KCl. For mean spike parameters see Table S1 (Supplementary).

**Figure 4.**

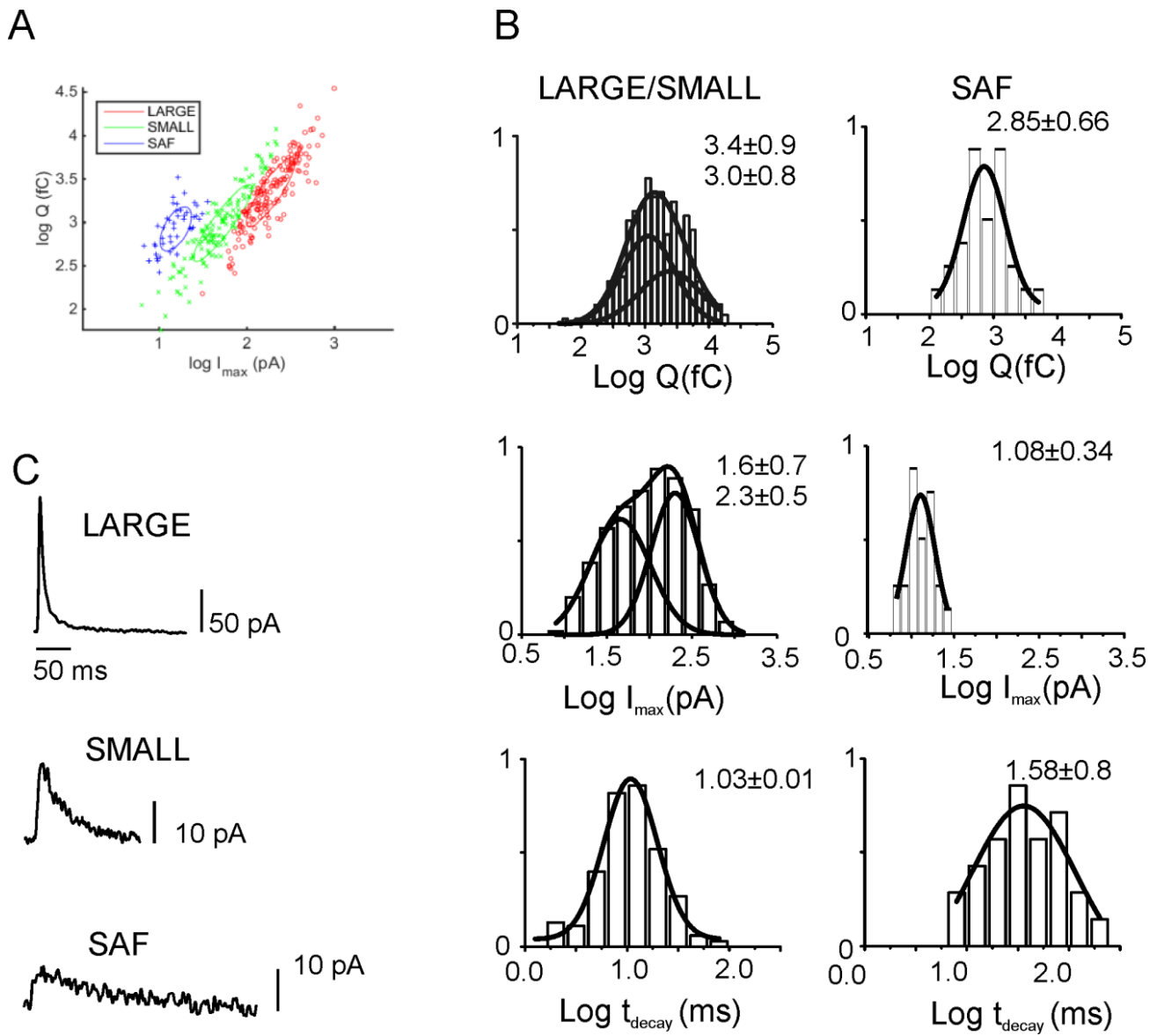

**Figure 4. Different modes of exocytosis in isolated BCCs**

A) 2D Gaussian mixture analysis of Log($Q$(fC)) values respect to Log($I_{max}$(pA)). Data were obtained from 4 cells (347 spikes). The cluster centers, the slopes, the correlation coefficients and the cluster percentage contribution to the total exocytosis are given in the text, while variances are represented by solid lines. B) Log-transformed distributions for $Q$, $I_{max}$ and $t_{decay}$ values. C) Representative events classified as large, small and SAF (see text for definitions).

**Figure 5.**

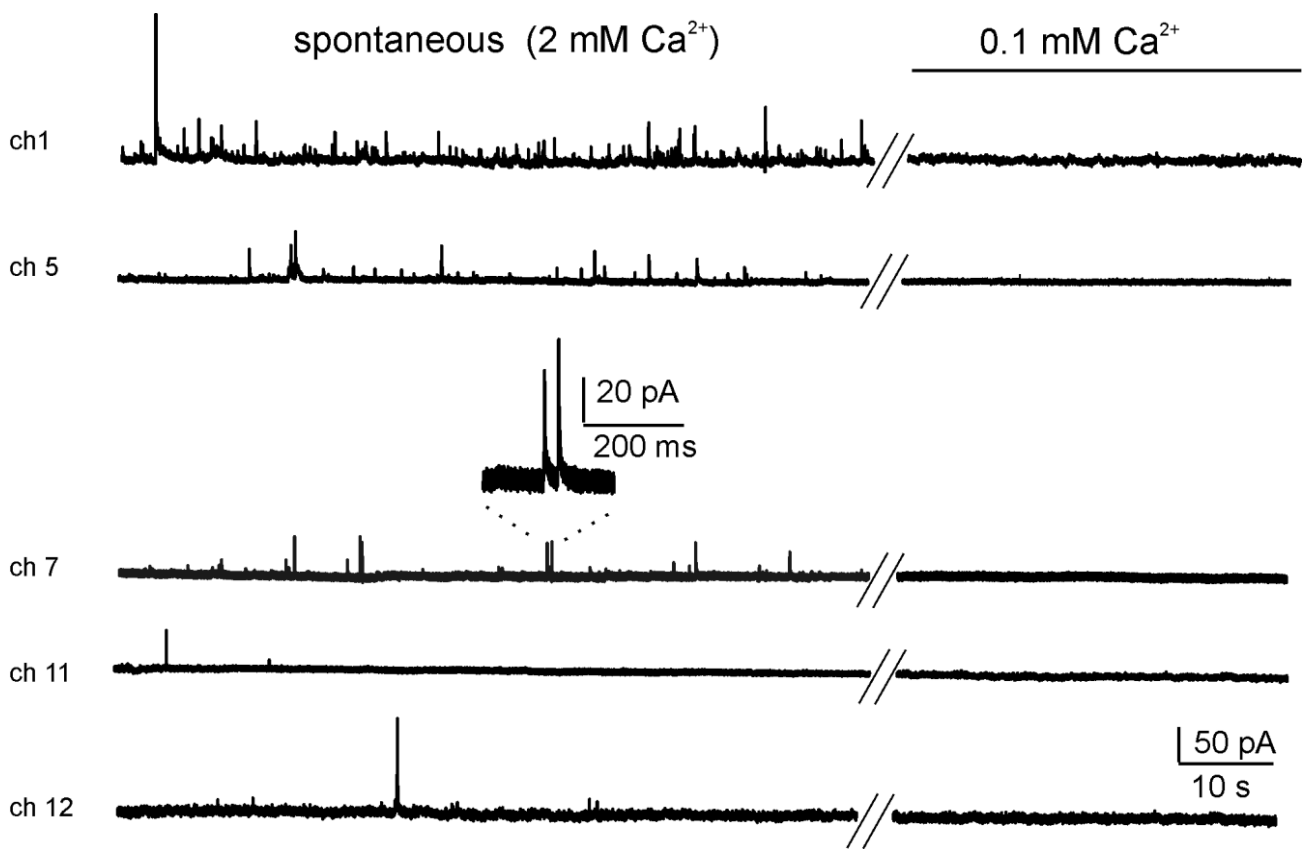

**Figure 5. Spontaneous exocytosis from a mouse adrenal slice placed on top of the 4×4 μG-SCD MEA**

Amperometric spikes recorded from 5 different channels of the microchip in 2 mM external $Ca^{2+}$ (left) and their suppression by adding a low $Ca^{2+}$ containing solution (right).